\documentclass{llncs}
\usepackage{graphicx}
\usepackage{array}
\usepackage{url}
\usepackage{times}
\usepackage{hyperref}
\usepackage{xspace}
\usepackage{pifont}
\usepackage{xcolor}
\usepackage{multirow,makecell}
\usepackage{booktabs}
\usepackage{hyperref}
\authorrunning{H. Althebeiti and D. Mohaisen}
\usepackage{setspace}
	
\usepackage{wrapfig,lipsum,booktabs}
\usepackage{xcolor}
\usepackage{soul}

\newcommand{\hlc}[2][yellow]{{%
    \colorlet{foo}{#1}%
    \sethlcolor{foo}\hl{#2}}%
}

\usepackage{esvect}
\usepackage{xspace}
\newcommand{\BfPara}[1]{{\noindent\bf#1.}\xspace}
\hypersetup{
	plainpages=false,
	colorlinks,
	urlcolor=blue!50!black,
	linkcolor=red!50!black,
	citecolor=green!50!black,
	bookmarksnumbered
}

\usepackage{makecell}
\usepackage{array}
\newcolumntype{L}[1]{>{\raggedright\let\newline\\\arraybackslash\hspace{0pt}}m{#1}}
\newcolumntype{C}[1]{>{\centering\let\newline\\\arraybackslash\hspace{0pt}}m{#1}}
\newcolumntype{R}[1]{>{\raggedleft\let\newline\\\arraybackslash\hspace{0pt}}m{#1}}

\begin{document}
\title{Enriching Vulnerability Reports Through Automated and Augmented Description Summarization}

\author{Hattan Althebeiti \and David Mohaisen}
\institute{University of Central Florida, Orlando, USA\\
\email{\{hattan.althebeiti,mohaisen\}@ucf.edu}}
% %
\maketitle              % typeset the header of the contribution
\begin{abstract}
Security incidents and data breaches are increasing rapidly, and only a fraction of them is being reported.
Public vulnerability databases, e.g., national vulnerability database (NVD) and common vulnerability and exposure (CVE), have been leading the effort in documenting vulnerabilities and sharing them to aid defenses.
Both are known for many issues, including brief vulnerability descriptions.
Those descriptions play an important role in communicating the vulnerability information to security analysts in order to develop the appropriate countermeasure.
Many resources provide additional information about vulnerabilities, however, they are not utilized to boost public repositories. 
In this paper, we devise a pipeline to augment vulnerability description through third party reference (hyperlink) scrapping.
To normalize the description, we build a natural language summarization pipeline utilizing a pretrained language model that is fine-tuned using labeled instances and evaluate its performance against both human evaluation (golden standard) and computational metrics, showing initial promising results in terms of summary fluency, completeness, correctness, and understanding.% consolidate additional resources provided by a third party and extract useful information to enhance vulnerability's description presented by NVD. We leverage natural language processing to build a more comprehensive summary that enrich the original description without losing any critical information. Our initial result show that our methodology is heavily dependent on the quality of the collected reports and their semantic similarities with the original description. It also shows that its is important to ensure variety between the collected reports. Our work also aims toward constructing knowledge specific dataset from available resources that can be utilized by pretrained models.

\keywords{Vulnerability \and NVD \and CVE  \and Natural Language Processing \and Summarization  \and Sentence Encoder \and Transformer.}
\end{abstract}
%
%
%
\if0
introduction: 

* dataset:
    - describe the source of the dataset: NVD, and why you use NVD for that. 
    - the set of steps that you have done to obtain the dataset: because you are only interested in recent vulnerabilities you limit your attention to vulnerabilities reported from 2019 to 2021, although our approach can be applied to anything. Then describe how 
    - heuristic to reduce the dataset into the relevant parts. Relevance is done based on semantics, length, etc. If you are using a representation for the similarity (semantic-preserving similarity) then describe those encoders.  
    - characteristics of the final dataset you use. 
    
    Methodology:
    **** PIPELINE:
    source dataset (augmented) -> target summary (more of a style).
    1) we wanted to enrich the target summarization so that it is inclusive of more details, thanks to the power of the pretrained model and its ability to generate new summary content. 
    2) normalization: we believe in the power of the model into normalizing different input source text into unified target summary.
    
    Source (s) -> tokenizing (t) -> encoding (e) T5 (learning network) -> intermediate output (i) -> decoded output (d). 
    
    Description of each step of the pipeline. 
    
    Results:
    * input data
    * out dataset
    * summary quality evaulation
    * etc. 
    
Conclusion

\fi

\section{Introduction}
%\subsection{A Subsection Sample}
Vulnerabilities are weaknesses in systems that render them exposed to any threat or exploitation. They are prevalent in software and are constantly being discovered and patched. However, given the rapid development in technologies, discovering a vulnerability and developing a mitigation technique become challenging. Moreover, documenting vulnerabilities and keeping track of their development become cumbersome. 

The common vulnerability and exposure CVE managed by MITRE and the National vulnerability database NVD managed by NIST are two key resources for reporting and sharing vulnerabilities. The content of each resource may differ slightly according to~\cite{ref_proc1}, but they are mostly synchronized and any update to the CVE should appear eventually in the NVD. However, NVD/CVE descriptions have several shortcomings. For example, the description might be incomplete, outdated or even contain inaccurate information which could delay the development and deployment of patches. In 2017 Risk Based Security also known as VulbDB reported 7,900 more vulnerabilities than what was reported by CVE ~\cite{web_ref,web_ref2}. Another concern with the existing framework is that the description provided for vulnerabilities is often incomplete, brief, or does not carry sufficient contextual information~\cite{abs-2006-15074,AnwarKNM18}. 

To address some of these gaps, this work focuses on the linguistic aspects of vulnerability description and attempts to improve them by formulating the problem as a summarization task over augmented initial description. We exploit the existence of third party reports associated with vulnerabilities, which include more detailed information about the vulnerabilities that goes beyond the basic description in the CVE. Therefore, we leverage these additional resources employing a natural language processing (NLP) pipeline towards that goal, providing informative summaries that cover more details and perform well on both computational and human metrics.

\BfPara{Contributions} The main contributions of this work are as follows. (1) we present a pipeline that enriches the description of vulnerabilities by considering semantically similar contents from various third party resources (reference URLs). (2) In order to normalize the enriched description and alleviate some of the drawbacks of the augmentation (e.g., redundancy and repetition, largely variable length of description), we build an NLP pipeline that exploits advances in representation, pretrained language models that are fine-tuned using the original (short description) as a label, and generate semantically similar summaries of vulnerabilities. (3) We evaluate the performance of the proposed NLP pipeline on NVD, a popular vulnerability database, with both computational and human metric evaluations. 

\section{Related Work}
Vulnerabilities are constantly being exploited due to the wide spread of malware and viruses along with the improper deployment of countermeasures or missing security updates.
Mohaisen et al.~\cite{mohaisen2015amal} proposed AMAL, an automated system to analyze and classify malware based on its behaviour. 
AMAL is composed of two components AutoMal and MaLabel. 
AutoMal collect information about malware samples based on their behaviours for monitoring and profiling. 
On the other hand, MaLabel utilizes the artifacts generated by AutoMal to build a feature vector representation for malware samples.  
Moreover, MaLabel builds multiple classifiers to classify unlabeled malware samples and to cluster them into separate groups such that each group have malware samples with similar profiles. 

Public repositories provide comprehensive information about vulnerabilities, however, they still suffer from quality and consistency issues as demonstrated in previous works~\cite{abs-2006-15074,ref_proc1}. 
Anwar et al.~\cite{abs-2006-15074} have identified and quantified multiple quality issues with the NVD and addressed their implications and ramifications.
The authors present a method for each matter to remedy the discovered deficiency and improve the NVD.
Similarly, Anwar et al.~\cite{anwar2020measuring} studied the impact of vulnerability disclosure on the stock market and how it affects different industries. 
They were able to cluster industries into three categories based on the vulnerabilities impact on the vendor's return.

Limited prior works studied different characteristics of vulnerabilities and used NLP based-approach on the task, although NLP has been utilized extensively for other security and privacy applications.
Alabduljabbar et al.~\cite{alabduljabbar2021automated} conducted a comprehensive study to classify privacy policies established by a third party.
A pipeline was developed to classify text segments into a high-level category that correspond to the content of that segment.
Likewise, Alabduljabbar et al.~\cite{alabduljabbar2022measuring} used NLP to conduct a comparative analysis of privacy policies presented by free and premium content websites.
The study highlighted that premium content websites are more transparent in terms of reporting their practices with respect to data collection and tracking.

Dong et al.~\cite{ref_proc1} built VIEM, a system to capture inconsistency between CVE/NVD and third party reports utilizing Named Entity Recognition model (NER) and a Relation Extractor model (RE).
The NER is responsible for identifying the name and version of vulnerable software based on their semantics and structure within the description and label each of them accordingly.
The RE component utilizes the the identified labels and pairs the appropriate software name and version to predict which software is vulnerable.

Other research focused on studying the relationship between CVE and Common Attack Pattern Enumeration and Classification (CAPEC) and if it is possible to trace a CVE description to a particular CAPEC using NLP as in Kanakogi et al.~\cite{ref_article1}. 
Similarly, Kanakogi et al.~\cite{ref_proc3} tested a new method for the same task but using Doc2Vec. 
Wareus and Hell~\cite{ref_lncs1} proposed a method to automatically assigns Common Platform Enumeration (CPE) to a CVEs from their description using NLP.  

\BfPara{This work} We propose a pipeline for enriching the vulnerability description, and a pipeline for normalizing  description through summarization and associated evaluation.

\section{Dataset: Baseline and Data Augmentation}
\BfPara{Data Source and Scraping} Our data source is NVD because it is a well-known standard accepted across the globe, in both industry and academia, with many strengths: (1) detailed structured information, including the severity score and publication date, (2) human-readable descriptions, (3) capabilities for reanalysis with updated information, and (4) powerful API for vulnerability information retrieval. 

In our data collection, we limit our timeframe to vulnerabilities reported between 2019 and 2021 (inclusive). Based on our analysis, CVEs reported before 2019 do not include sufficient hyperlinks with additional text, which is our main source for augmentation. 
We list all the vulnerabilities reported in this period, and scrap them. For each vulnerability, we  scrap the URLs pointing to the NVD page that hosts a particular vulnerability.
As a result, we obtain 35,657 vulnerabilities with their unique URLs. Second, we iterate through every URL various data elements. 
After retrieving the URL, we scrap the description and the hyperlinks for that vulnerability. 

\BfPara{Description Augmentation}
To augment the description, we iterate through the scrapped hyperlinks.
Each hyperlink directs us to a page hosted by a third party, which could be an official page belonging to the vendor or the developer or an unofficial page; e.g., GitHub issue tracking page.
We scrape every paragraph tag in each page separately and apply various preprocessing steps to the extracted paragraph to clean up the text. 
This preprocessing includes removing web links, special characters, white redundant spaces, phone numbers, and email addresses. 
We also check the length of the paragraph and ensure it is more than 20 words after preprocessing. We conjecture that paragraphs shorter than 20 words will not contribute to our goal. 

After cleaning the text and verifying the length, we use a sentence encoder to encode the semantic for the extracted paragraph and the scrapped description into low dimensional vector representations (more in \textsection\ref{sec:se}).
To determine the similarity between the vectorized representations, we use the cosine similarity which yields a value between -1 and 1. For example, let the vector representation of the extracted paragraph be $\mathbf{v}_p$ and that of the description be $\mathbf{v}_d$, the cosine similarity is defined as:
\begin{equation}
    \cos(\mathbf{v}_p, \mathbf{v}_d) = \frac{ \vv{v_p}\cdot \vv{v_d} }{||\vv{v_p}||\hspace{0.1cm}||\vv{v_d}||}
\end{equation}

If $\cos(\mathbf{v}_p, \mathbf{v}_d)$ exceeds a predefined threshold, we add/augment the paragraph as the input text and the description as the summary text.  
This process is repeated with every paragraph contained within a page. 
We repeat this step for every hyperlink by extracting the page, associated paragraph tags, applying preprocessing, encoding semantic and measure the similarity with the description.  We note that some vulnerabilities may not be added to our dataset; e.g., if the vulnerability did not have any hyperlinks or if its associated hyperlinks did not include any paragraph that meets the predefined threshold. We repeat the process for each URL until we cover all the URLs, upon which dataset is ready to be presented to the model.

%However, we note that adding the augmented text and the description to the dataset is the last step, meaning that after we go through every hyperlink and augment all paragraphs that are similar to the description, we add the entry to the dataset.

%Upon going through all hyperlinks and adding an entry to the dataset, we move to the next URL in our outer loop, which is a new vulnerability with new hyperlinks. 

\begin{figure}[t]
    \centering
    \includegraphics[width=1.02\textwidth]{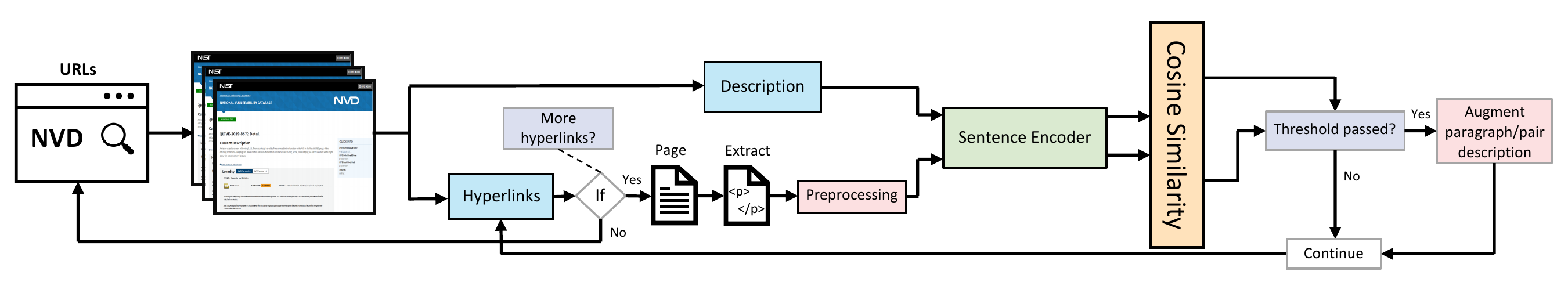}\vspace{-3mm}
    \caption{Data collection pipeline}
    \label{figure:Pipeline}\vspace{-5mm}
\end{figure}
 
\autoref{figure:Pipeline} shows our pipeline. The choice of a sentence encoder will affect the dataset because the inclusion of a paragraph is based on the similarity score between the vectorized representation of the description and paragraph encoded by the sentence encoder. 
To enhance our experiments and provide a better insight into different encoders and summarization models, we use two sentence encoder choices: Universal Sentence Encoder (USE) and MPNet sentence encoder.
%We note that the outcome of the encoder may influence the eventual dataset realized through this pipeline, and use that diversity in outcome to highlight the better model through fine-tuning. 
In our analysis, we use the best performing encoder with respect to the end-goal outcome of our summarization task. 

Per \autoref{figure:Pipeline}, the similarity score must exceed a predefined threshold.
From our preliminary assessment of  the two encoders, we found that USE is more accurate (sensitive) than MPNet in terms of the similarity score representation, meaning that when the description and the paragraph are (semantically) similar to one another, USE produces a higher score than MPNet and vice versa. 
Considering this insight, we set different threshold for each encoder.
Namely, we set the similarity score for USE to be between 0.60 and 0.90, since the encoder is accurate.
On the other hand, since MPNet is less accurate (sensitive) than USE, we enforce a more restrictive threshold and set it between 0.70 and 0.90. 
We excluded paragraphs with a similarity score above 0.90 because we found those paragraphs to be almost identical to the description, thus adding them would not serve the main purpose of enriching the description. 
Those values were picked as part of our assessment over the two encoders using a small set of vulnerabilities and following the procedure explained above. 

Some hyperlinks analysis took extremely long time. Upon examining  the content of those pages, we found that they contain a history of the software vulnerability with updates, e.g., over 20,000 paragraph tags in some cases. Moreover, most of them were not considered by the sentence encoder because they do not meet the threshold.
As such, we consider the first 100 paragraph tag in each hyperlink to speed up the process. We justify this heuristic by noting  that most pages contain the related textual information at the beginning with subsequent paragraphs being reiteration of information that is already mentioned earlier. 
Finally, to only limit our collection to authentic descriptions, we consider hyperlinks with valid SSL certificate.

Additionally, we curated a third dataset using both and enforcing multiple thresholds on the similarity criterion. 
For that, we used the same the threshold for the MPNet as before, and  relaxed the threshold for USE to 0.50 to  relax an imposed restrictive setting by possibly excluding otherwise qualified candidate paragraphs. 

\begin{wraptable}{r}{3.7cm}
\centering\vspace{-9mm}\renewcommand{\arraystretch}{0.85}
\caption{Datasets}\vspace{2mm}
\begin{tabular}{|l|l|r|}
  \Xhline{2\arrayrulewidth}
    \# CVEs & Encoders & Vuln. \\
    \Xhline{1\arrayrulewidth}
    \multirow{3}{*}{35,657} & USE & 9,955  \\
    \cline{2-3}
    & MPNet & 8,664  \\
    \cline{2-3}
    & Both & 10,766 \\
    \Xhline{2\arrayrulewidth}
  \end{tabular}\vspace{-2mm}
\label{table:datasets}\vspace{-5mm}
\end{wraptable}
Given the differences between the two encoders, we consider a paragraph to be similar if the difference between the two similarity scores is at most 0.20; otherwise we consider them dissimilar and discard the paragraph. Here, we favored the consistency between the two encoding techniques to conceptually alleviate the discrepancy presented from using the two different encoders. 
Table \ref{table:datasets} shows the datasets. 
In the next section, we elaborate about the encoders in more detail.

\section{Methodology and Building Blocks}

\subsection{Sentence Encoders}\label{sec:se}

Among the multiple tried encoders over multiple CVEs along with their similar paragraphs, we found that the best encoders for our task are the Universal Sentence Encoder (USE)~\cite{ref_proc9} and MPNet sentence encoder \cite{ref_proc8}, which we explain in the following. 

\BfPara{Universal Sentence Encoder}  \label{USE}
Two architectures are proposed for USE. 
The first is a transformer-based model which uses a transformer architecture to compute context aware representation of the words while preserving words’ positions, followed by embeddings used to compute fixed length sentence encoding using element-wise sum at each word position.
The downside of this architecture is its time and space complexities, i.e., it takes $O(n^2)$ and is proportional in space to the sentence length. 
The second architecture is much simpler and uses a Deep Averaging Network (DAN).
It computes a sentence initial embedding by averaging words with bi-gram embeddings and passes this embedding through a feed forward network to produce the final embedding. 
Unlike the transformer architecture, DAN's time complexity is $O(n)$ and its space is constant with respect to the length of the sentence.
The trade-off in choosing among those two architectures is between the high accuracy with intensive computation achieved by the transformer architecture versus the efficient inference and computation with a reduced accuracy achieved by the DAN architecture. 
Given our problem's characteristics, we decided to use the DAN architecture because  (1) our data will be scraped, and its length may vary widely, and (2) our data is domain-specific and is limited in its linguistic scope. 
We conjecture DAN will produce accurate embedding since the vocabulary size is limited (i.e., small). 
Finally, considering that we have over 35,000 Vulnerabilities, where each has multiple hyperlinks to be scraped, the scalability benefit of DAN outweighs the high accuracy of the transformer-based architecture.

\BfPara{MPNet}
The second technique we utilize is MPNet. MPNet is a model that leverages the advantages presented in two famous pretrained models: BERT \cite{ref_proc5} and XLNET \cite{ref_proc7}. 
BERT uses a masked language modeling objective, which masks 15\% of the tokens and the model is trained to predict them. 
The downside of BERT is that it does not consider the dependency between the masked tokens. 
On the other hand, XLNET retains the autoregressive modeling by presenting permuted language modeling objective in which each token within a sequence considers the permutations of the previous tokens in the sequence but not after it. 
However, this causes position discrepancy between the pretraining and fine-tuning. MPNet unifies the objectives of the two models by considering dependency among predicted tokens and considering all tokens’ positions to solve the position discrepancy. 
Moreover, MPNet sentence transformer is built by fine-tuning MPNet on 1 billion sentence-pair dataset and uses contrastive learning objective. 
Given a sentence from the pair, the model tries to predict which other sentence it was paired with.
This is done by computing the cosine similarity with every other sentence in the batch and then using the cross-entropy loss with respect to the true pair.
In the next section we explain the pipeline for our summarization models. 

\subsection{Pretrained Models}\label{sec:pretrained}
\begin{figure}
    \centering\vspace{-4mm}
    \includegraphics[width=0.85\textwidth]{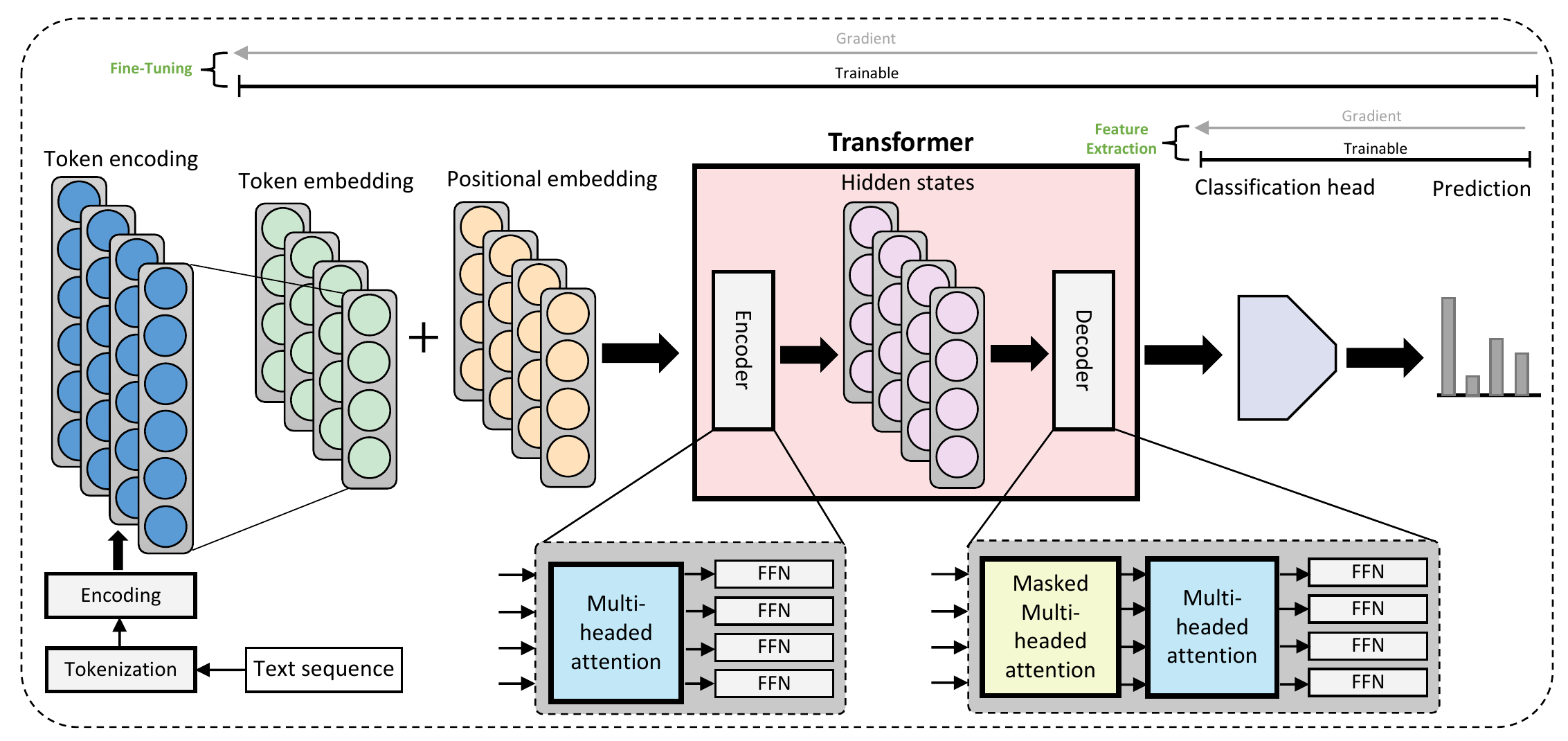}\vspace{-4mm}
    \caption{Our summarization pipeline}
    \label{figure:Transformer_Pipeline}\vspace{-6mm}
\end{figure}

The goal of this work is to use pretrained models and fine-tune them on our datasets for vulnerability summarization and description enrichment. 
The pretrained models inherit the architecture of the original transformer \cite{ref_proc10} with some adjustments to the weights depending on the task it is performing. 
The transformer itself constitutes of two major components: an encoder and a decoder. 
The encoder’s role is to build a representation for the input sequence that captures the dependencies between tokens in parallel without losing positional information of those tokens. 
The transformer relies on the attention mechanism to capture interdependency within a sequence, which provides a context-aware representation for each token. 
The decoder’s role is to use the built representation and map it to a probability distribution over the entire vocabulary to predict the next word. 
Figure \ref{figure:Transformer_Pipeline} shows  the pipeline of a the encoder-decoder transformer from the beginning of inputting the raw text to the prediction (decoded into utterances for sequences generation; i.e., summarization). 

The original transformer was developed and is intended for machine translation, although generalized to other tasks with remarkable results.  
We note that most modern pretrained models use a transformer architecture that depends on an encoder only; e.g., BERT \cite{ref_proc5}, a decoder only; e.g., GPT (Generative Pre-trained Transformer)~\cite{ref_proc12}, or both. 
Each architecture has its own advantages, which allows it to excel in specific tasks. 
The summarization task, for example, can be modeled as a seq2seq task where the model takes an input (long text) and outputs the summary, which naturally makes a model that constitutes of encoder and decoder ideal for its design. 
In the NLP literature, the most performant models for summarization are BART \cite{ref_proc11}, T5 \cite{ref_proc6}, and Pegasus \cite{ref_proc13}, with BART and T5 being more widely used. BART is a denoising autoencoder for pretraining seq2seq with an encoder-decoder architecture.
The idea of BART is to use a noising function to corrupt the text and train the model to reconstruct the original (uncorrupted) text. In contrast, T5 uses a masked language modelling objective like BERT for training. Instead of masking a token, T5 masks a span of the original text as its corruption strategy.
The length of the span does not influence the model performance unless too many tokens are within that span.
%The original paper did an extensive experimentation to find the best set of parameters. 
Moreover, T5 attempts to define a framework for many NLP tasks by adding a prefix that identifies the task it tries to learn. 
Therefore, one model can support multiple tasks by defining those prefixes in the training data and adding those prefixes to a sample allows the model to predict for the task associated with that prefix. 
%Finally, T5 follows the original transformer architecture with a slight modification among its components. 

\subsection{Pipeline}\label{sec:pipeline}
Next, we discuss the pipeline depicted in \ref{figure:Transformer_Pipeline} in more details. The major steps of our pipeline are tokenization of the input text sequence (description), encoding, token embedding, positional embedding, encoding-decoding (utilizing a fine-tuned pretrained language model), and prediction. Those steps are elaborated in the following.

\BfPara{Tokenization} The first step for most models is tokenization, which includes breaking text into individual independent entities and encoding them into numerical representation. 
Tokenization could be applied at the word or character level. With word tokenization, we will end up with a large vocabulary size that will affect the dimensionality of the word embedding. 
To address the dimensionality issue, it is common to limit the size to the most common 100,000 words in a corpus and encode all unknown words as $<$UNK$>$.
However, most words morphemes will be encoded as unknown although they possess very similar meaning to their root.
Similarly, character embedding dominant limitation is losing the linguistic structure and considering a text as a stream of characters. A third type is the subword tokenization, which alleviates the drawbacks of the two aforementioned tokenization granularities. Subword tokenization splits rare words into a meaningful unit which helps the model to handle complex words and associate their embedding with similar words.
This allows the model to associate singular with plural and relate different morphemes to their root.
BART uses Byte-Pair Encoding (BPE)~\cite{ref_proc15} and T5 uses SentencePiece~\cite{kudo2018subword,kudo2018sentencepiece} which are both subword tokenizer. 

\BfPara{Token Encoding} The tokenized text is transformed into numerical representation using one-hot encoding with a size equal to the vocabulary size; e.g., 20k-200k tokens.

\BfPara{Token Embedding} The token encodings are then projected into lower dimensional space that captures the characteristics of each word in a token embedding. 
However, for a pretrained model this and the previous step are already done and the token embedding is already computed during the model training. Those two steps are required if we plan to build our own transformer from the bottom-up.  
In practice, each token will be represented by an id that identifies it with respect to the model.% as shown in Figure \ref{figure:Tokenization}.
%However, it is important to shed some lights on two additional yet significant concepts in that regard which are attention mask and padding. 

% \begin{figure}[t]
%     \centering
%     \includegraphics[width=1.02\textwidth]{Tokenization.pdf}
%     \caption{Tokenizating and embedding}
%     \label{figure:Tokenization}\vspace{-5mm}
% \end{figure}

Each text consists of tokenized words and each token is represented by an input id. 
To increase the efficiency of the model, we create a batch of multiple text before feeding the text into the transformer. 
However, to create a batch, we must ensure that all texts have the same size as the longest text in that batch.
For that padding is used to pad short text to meet the length  requirements by adding id ‘0’ to the text sequence. Moreover, the attention mask informs the model to ignore those padding during encoding by assigning 1 to tokens that are part of the original sequence and 0 for padding.
Finally, the batch of texts (with attention masks) is passed to the transformer block.
Each model has some reserved ids that are used for a specific purpose.
%For example, we see that in all three texts that start with id ‘0’ and end with ‘2’, these are the ids used by BART to indicate the beginning and ending of a text. 
%Each model has its own specific IDs which is why it is important to use the tokenizer associated with the model being used. Another important parameter is truncation which instruct the model to truncate a sequence after maximum number of tokens. 

\BfPara{Positional Embedding} The transformer uses the attention mechanism to capture the contextual interdependence between words. 
However, this method is oblivious to the words’ positions, and we need a way to inject this information into the word embedding.
As with tokenizer, each model has its own way of including this information.
BART uses the same method used in the original transformer where a simple sinusoidal function is used to create a positional embedding for each token within a sequence.
On the other hand, T5 uses a more sophisticated approach, called the Relative Positional Encoding (RPE), which uses a multi-headed attention to encode the relative positions between tokens.
The intuition behind RPE stems from the fact that what is most important is the surrounding words rather than its exact position, and that is how RPE computes the positional embedding. 
The token embedding, and the positional embedding are added together to build the final embedding that will be fed into the transformer.

\BfPara{Transformer} 
This step consists of an encoder an a decoder.  The encoder uses a multi-headed attention to build a representation that captures the contextual interdependence relationship between tokens.
The encoder uses several layers of self-attention to compute how much attention should be paid by every token with respect to other tokens to build the final numerical representation.
Modern transformers use the scaled dot product attention which utilizes a query, key, and value computed for each token to produce the attention score for every token with respect to other tokens in the sequence.
A simple intuition behind applying several attention layers (heads) is that each head may focus on one aspect of attention, while others may capture a different similarity.
By concatenating the output of all heads, however, we obtain a more powerful representation that resembles that sequence.
The feed forward network receives every token embedding from the multi-headed attention and processes it independently to produce its final embedding which is referred to as the hidden states.

As the encoder outputs a representation of the input sequence, the decoder’s objective is to leverage the hidden states to generate the target words.  We note that summarization requires text generation to generate the next token in an autoregressive fashion.
As such, the generation procedure’s objective is to predict the next token given the previous tokens.
This can be achieved using the chain rule to factorize the conditional probabilities as
\begin{equation}
P(x^({^t}{^+}{^1}{^)}|x^({^t}{^)},...,x^({^1}{^)}) = \prod_{t=1}^{T} P(x^({^t}{^+}{^1}{^)}|x^({^t}{^)},...,x^({^1}{^)})
\end{equation}
A numerical instability results from the product of the multiple probabilities as they become smaller. Thus, it is common to use the log of the conditional probability to obtain a sum, as

\begin{equation}
\log(P(x^({^t}{^+}{^1}{^)}|x^({^t}{^)},...,x^({^1}{^)})) = \sum_{t=1}^{T} \log(P(x^({^t}{^+}{^1}{^)}|x^({^t}{^)},...,x^({^1}{^)}))
\end{equation}
 
From this objective, there are various methods to select the next token through decoding with two aspects to consider. (1) The decoding method is done iteratively, where the next token is chosen based on the sequence at each time step.
%The selected token is then appended to the sequence and the new sequence is used by the decoding method to generate the next token and so on, until it reaches the maximum length or end of sequence (EOS).
(2) It is important to emphasize certain characteristics of the selected word; e.g., in summarization we care about the quality of the decoded sequence, compared to storytelling or open domain conversation where care more about the diversity when generating the next token.

\BfPara{Decoding} In this work, the beam search is used as decoder, since  summarization emphasizes factual or real information in the text.
This method is parameterized by the number of beams, which defines the number of the most probable next tokens to be considered in the generated sequence and keep track of the associated sequences by extending a partial hypothesis to include the next set of probable tokens to be appended to the sequence until it reaches the end of sequence.
The sequences are then ranked based on their log probabilities, and the sequence with the highest probability is chosen.
It is important to ensure that at each time step, the decoder is conditioned on the current token and the past output only.
% This is the role of the masked multi-headed attention component.
This step is crucial to assure the model does not cheat by accessing future tokens.
While the transformer architecture is task-independent, the classification head is task-specific, and we use a linear layer that produces a logit followed by a softmax layer to produce a probability distribution for decoding. 

\BfPara{Operational Considerations} Transformers are typically deployed in one of two setting. (1) As a feature extractor, where we compute the hidden states for each word embedding, the model parameters are frozen, and we only train the classification head on our task. Training using this method is fast and suitable in the absence of resources to fine tune the whole model. (2) As a fine-tuning setting, where all the model trainable parameters are fine-tuned for our task. This setting requires time and computational resources depending on the model size. 
In our case we use BART and T5 for fine-tuning and since BART has a smaller number of parameters, its fine-tuning is faster.

\section{Evaluations}

\BfPara{Statistical Analysis}
After assembling the three datasets, we picked the dataset produced by both encoders, given that it is the largest, for statistical analysis (the results with other datasets are omitted for the lack of space). The goal of this analysis is to obtain a better insight over the dataset language characteristics. From this analysis, we found the number of tokens of most augmented descriptions falls below 1000 tokens, in contrast to the original summary which is below 200 tokens for the majority of vulnerabilities. 
Therefore, we set the threshold for the augmented description and the summary to be 1000 and 250 tokens, respectively, in our pipeline. 
%Figure \ref{figure:Word_dist} shows the distribution of the tokens across both features, summary and augmented text. 
% \begin{figure}
%     \centering
%     \includegraphics[width=0.99\textwidth]{Word_dist.pdf}
%     \caption{Tokens Distribution}
%     \label{figure:Word_dist}
% \end{figure}

%During the fine-tuning the models, we truncate any sequence beyond those limits to speed up training and reduce computational resources.

% \begin{table}[h]
% \centering
% \begin{tabular}{m{2cm}m{2cm}m{2cm}m{2cm}m{2.2cm}}
%   \Xhline{2\arrayrulewidth}
%     Feature & Measure & $w$ & $c$ & $s$ \\
%     \Xhline{1\arrayrulewidth}
%     \multirow{2}{*}{Text} & mean & 481.28 & 2838.01 & 43.73 \\
%     & std & 2086.04 & 12370.67 & 184.61 \\
%     \Xhline{1\arrayrulewidth}
%     \multirow{2}{*}{Summary}& mean & 49.60 & 279.92 & 7.49 \\
%     & std & 31.89 & 186.35 & 5.32 \\
%     \Xhline{2\arrayrulewidth}
%   \end{tabular}
% \caption{Summary statistics over augmented text and summary: word ($w$), character ($c$), and sentence ($s$).}
% \label{table:sum_stats}\vspace{-8mm}
% \end{table}

We collect the word, character, and sentence count of the augmented and original summary and found a significant difference between them (e.g., (mean, standard deviation) for word, character, and sentence in both cases are: (48, 2086) vs (49, 31), (2939, 12370) vs (279, 186), and (43, 184) vs (7, 5.32). This highlights the need for a summarization to normalize the augmented description.% summarization dataset which was confirmed after training our models. 

% \begin{figure}[h]
%     \centering
%     \includegraphics[width=0.76\textwidth]{Top_Named_Entities.pdf}
%     \caption{Named Entity Recognition}
%     \label{figure:NER}
% \end{figure}

Next, we perform named entity recognition to understand which entities were presented across the summary because this is our target in the dataset.
%%this is not good. If the entity recognition is done for the original summary, then you also need to do that for the augmented description and the associated summary. 
We found the following frequent named entities: (XSS, \hlc[pink]{799}), (N/AC	\hlc[pink]{523}), (IBM X-Force ID,	\hlc[pink]{463}), (N/S,	\hlc[pink]{343}), (Cisco,	\hlc[pink]{336}), (SQL,	\hlc[pink]{334}), (Server,	\hlc[pink]{315}), (JavaScript,	\hlc[pink]{267}), (WordPress,	\hlc[pink]{264}), (Jenkins,	\hlc[pink]{240}), (IBM, \hlc[pink]{237}), (Firefox, \hlc[pink]{200}), (Java, \hlc[pink]{187}), (VirtualBox, 	\hlc[pink]{174}), (PHP,	\hlc[pink]{164}), (Java SE,	\hlc[pink]{150}), and (Android,	\hlc[pink]{148}). The common names include organizations, e.g., Cisco and IBM, technologies, e.g., JavaScript, and PHP, or vulnerabilities, e.g., XSS.

We further analyze the most frequent trigram across the dataset. We found that the description trigrams are meaningful, and form the basis for a good summary, in contrast to the augmented text trigrams that, in general, do not present useful information and appear to be uninformative.
This might be a result of augmenting repeated content, which highlights certain trigrams based on the frequency. Those initial results highlight the need for an additional summarization step.  

% \begin{figure*}
%     \centering
%     \includegraphics[width=1.0\textwidth]{Trigram.pdf}
%     \caption{Trigram}
%     \label{figure:trigram}
% \end{figure*}

% \section{Evaluation}
% Two evaluation methods are considered that constitute of several metrics.
% Computational metrics are used to judge the overlap between the prediction and the target.
% Human evaluation is done through the means of human metrics that cannot be measured algorithmically.
% Metrics such as fluency and understanding are best judged by an evaluator to get a better insight on the model effectiveness on generating a summary.

\BfPara{Experimental Settings}
We fine-tune both models using two different settings.
First, We split the dataset with \%10 reserved for testing. Then, we split the training set with \%10 reserved for validation.
Second, based on our preliminary analysis, we set 1000 and 250 tokens as the maximum lengths for augmented descriptions and new summary. 
\begin{table}\vspace{-8mm}
\centering\renewcommand{\arraystretch}{0.90}
\caption{Results after fine-tuning the models using different hyperparameters ({\bf R}ecall, {\bf P}recision, $b$=number of beams, $T$=text maximum limit, $B$=batch size)}\vspace{-1mm}
  \begin{tabular}{L{0.08\textwidth}R{0.1\textwidth}R{0.1\textwidth}R{0.1\textwidth}R{0.1\textwidth}R{0.1\textwidth}R{0.1\textwidth}rr}
  \toprule
    Model & R & P & F1 & $T$ & $b$ & $B$\\
  \hline
    \multirow{6}{*}{BART} & 0.51 & 0.50 & 0.49 & 1000 & 2 & 8 \\
    & 0.51 & 0.46 & 0.47 & 1000 & 5 & 8  \\
    & 0.52 & \textbf{0.52} & \textbf{0.51} & 500 & 2 & 8 \\
    & \textbf{0.53} & 0.50 & 0.50 & 500 & 5 & 8 \\
    & 0.50 & 0.51 & 0.49 & 500 & 2 & 4 \\
    & 0.51 & 0.49 & 0.49 & 500 & 5 & 4 \\
    \hline
    \multirow{4}{*}{T5} & 0.46 & 0.50 & 0.47 & 500 & 2 & 8 \\
    &  0.47 & 0.49 & 0.47 & 500 & 5 & 8  \\
    & 0.47 & \textbf{0.52} & \textbf{0.48} & 500 & 2 & 4  \\
    & \textbf{0.47} & 0.50 & 0.47 & 500 & 5 & 4  \\
    \hline
  \end{tabular}\vspace{-3mm}
  \label{table:comp-results}\vspace{-0.4cm}
\end{table} \vspace{0.5mm}

Finally, We set the batch size to 8 and the learning rate to  0.0001 based on various parameters (results omitted for the lack of space). We use beam search as our decoding method, with a beam size of 2. We also fix several parameters: length penalty to 8 (which encourages the model to produce longer summary if it is set to a value higher than 1), and the repetition penalty to 2 (which instructs the model whether to use words that have already been generated or not). Those values are chosen among various values for their best performance, as demonstrated in Table~\ref{table:comp-results}. 
As we stated earlier, we did extensive experimentation on the mixed dataset that uses both encoders and based on its result we experimented with other datasets.

\BfPara{Computational Metrics and Results}
ROUGE measures the matching n-gram between the prediction and the target. For our evaluation, we use ROUGE-1, which measures the overlapping unigram, and gives an approximation of the overlap based on individual words.
For ROGUE, we use three sub-metrics: recall, precision, and F1 score. The recall measures the number of matching n-gram between our generated summary and the target summary, normalized by the number of words in the target summary.
In contrast, the precision normalizes that quantity by the number of words in the generated summary.
Finally, F1 score is expressed as:
\begin{equation}
    F1{-}Score = 2\times \frac{precision \times{recall}}{precision + {recall}}
\end{equation}

\begin{wraptable}{r}{3.5cm}
\centering\vspace{-8mm}\renewcommand{\arraystretch}{0.85}
\caption{Models training $T_\ell$ and validation loss $V_\ell$ over different batch sizes ($B$)}\vspace{2mm}
\begin{tabular}{|l|r|r|r|}
  \Xhline{2\arrayrulewidth}
    Model & $T_\ell$ & $V_\ell$ & $B$ \\
    \Xhline{1\arrayrulewidth}
    \multirow{2}{*}{BART} & 0.42 & 0.46 & 8  \\
    \cline{2-4}
    & 0.32 & 0.46 & 4  \\
    \hline
    \multirow{2}{*}{T5}& 1.96 & 1.48 & 8 \\
    \cline{2-4}
    & 2.35 & 1.46 & 4 \\
    \Xhline{2\arrayrulewidth}
  \end{tabular}\vspace{-4mm}
\label{table:losses}
\end{wraptable}
Table \ref{table:comp-results} shows the ROUGE scores after fine-tuning BART and T5.
Multiple experiments have been conducted using different batch sizes, text limit, and number of beams.
As we can see in Table \ref{table:comp-results}, when the text limit has shrunk to 500 tokens for the augmented text, all metrics have improved.
We also can see that most metrics achieved better score with a smaller number of beams.
This is explained by the beam search decoding, as we increase the number of sequences by having a high number of beams, the risk introduced by considering the wrong sequence increases.

We tested BART with a batch size of 4 and with 500 tokens as the augmented description limit and it outperformed the model trained with 1000 as text limit. It is important to notice that as the number of beams increases, the time it takes the model to generate the summary increases.
Considering our initial results from BART and the resources demand for T5 as it is much larger, we decided to train it on text limited to 500 tokens.
However, the results did not align with BART. For example, we found that batch size of 4 did better than 8 across all three metrics for T5.
Moreover, we see that increasing the number of beams did not help.
We point out, however, that the validation loss varies between the two models as shown in the Table~\ref{table:losses}.
This shows that BART did better than T5 during training, which is why BART achieved better scores.

{\noindent\bf\em Summary comparison} We compare the target summary with the model generated summary using the same sentence encoders. We encode both summaries (original and new) using both encoders and measure the similarity between the target and the prediction. We found that most predictions are very close to the target with the mean of the distribution around a similarity of 0.75 (the figures are omitted for the lack of space). 
%In Figure \ref{figure:similarity_figure}, we see that most predictions are considered close to the target. 

% \begin{figure*}
%     \centering
%     \includegraphics[width=0.90\textwidth]{similarity_figure.pdf}
%     \caption{Similarity between model generated summary and the target through the different encoders}
%     \label{figure:similarity_figure}
% \end{figure*}

We report the computational metrics in \autoref{table:results-encoders}. 
Although the mixed dataset had more instances, the models trained on the separate datasets outperformed it.
This could be attributed to the restriction we relaxed for the USE encoder, which allows the pipeline to include more paragraphs. Moreover, since the two encoders use different architectures, using them together may have a negative effect on the curated dataset.
More experimentation might be needed to find the perfect threshold to use them both.

\begin{wraptable}{r}{5.5cm}
\centering\vspace{-11mm}\renewcommand{\arraystretch}{0.85}
  \caption{Results after fine-tuning the models using different single encoder ({\bf P}recision, {\bf R}ecall, $b$=beams, $B$=batch)}\vspace{2mm}
  \begin{tabular}{|l|c|c|c|c|c|c|}
  \Xhline{2\arrayrulewidth}
    Model & Encoder & R & P & F1 & $b$ & $B$\\
  \Xhline{2\arrayrulewidth}
    \multirow{2}{*}{BART} & USE & 0.61 & 0.60 & 0.59 & 2 & 8 \\
    \cline{2-7}
    & MPNet & 0.55 & 0.57 & 0.55 & 2 & 8  \\
    \Xhline{2\arrayrulewidth}
    \multirow{2}{*}{T5} & USE & 0.58 & 0.62 & 0.59  & 2 & 4 \\
    \cline{2-7}
    & MPNet & 0.53 & 0.59 & 0.54 & 2 & 4  \\
    \Xhline{2\arrayrulewidth}
  \end{tabular}
  \label{table:results-encoders}\vspace{-7mm}
\end{wraptable}
The models trained using USE dataset outperformed the MPNet dataset.
While the USE dataset is larger, we believe the results are better due to USE's accuracy in encoding text semantic.
It also prove that USE produces a reliable representation for long text.
We reiterate here that we used the DAN architecture for USE which is less accurate than the transformer architecture as we explained in section \ref{USE}.
Therefore, using the transformer architecture to build the dataset could generate a more accurate dataset that is likely to outperform the result in Table \ref{table:results-encoders}.

\BfPara{Human Metrics Results}
We consider four human metrics: fluency, correctness, completeness, and understanding.
All human metrics are graded on a scale between 1-3 in which 3 is the best grade and
1 is the worse in terms of the metric definition. 

{\em Fluency} measures the grammatical structure of the prediction and how coherent the semantics of the generated
summary. The {\em correctness} measures how accurate the model prediction is in terms of capturing the correct vulnerability details. 
The {\em completeness} measures how complete is the generated summary with respect to details in the original summary.
The {\em understanding} measures how easy it is to understand the generated summary. The human evaluation on the generated summary from both models is done over 100 randomly selected samples where the average is reported in Table \ref{table:HE}. 
\begin{wraptable}{r}{3.8cm}\vspace{-5mm}
\centering
\caption{Human evaluation: {\bf F}luency, {\bf C}o{\bf m}pleteness, {\bf C}o{\bf r}rectness, and {\bf U}nderstanding}\vspace{2mm}
\begin{tabular}{|l|r|r|r|r|}
  \Xhline{2\arrayrulewidth}
    Model & F & Cm & Cr & U \\
    \Xhline{1\arrayrulewidth}
    BART & 2.69 & 2.15 & 2.16 & 2.58  \\
    \Xhline{1\arrayrulewidth}
    T5 & 2.72 & 2.07 & 2.04 & 2.57  \\
    \Xhline{2\arrayrulewidth}
  \end{tabular}
\label{table:HE}\vspace{-6mm}
\end{wraptable}

After analyzing both models we observed similar behaviors.
We found that both models produce a fluent summary with very few exceptions.
Similarly, the generated summaries are mostly easy to follow and understand.
However, in some cases when the generated summaries are short, they do not convey much meaning and it becomes hard to understand the summaries.
In contrast, completeness and correctness suffered with both models.
We did not anticipate the models to perform well across those metrics because the dataset was not curated for detecting such features.
Moreover, the dataset is imbalanced with respect to its features in terms of augmented text length which we believe is the main reason for both models in missing those two metrics.
However, when the augmented text is of certain length, those two metrics achieve good results.
The human evaluation metrics are averaged and shown in table \ref{table:HE}.
We can see that both models are comparable in terms of human metrics when their generated summary is compared against the corresponding target.

\BfPara{Qualitative Results}  Both models experienced unpredictable behaviors by repeating some sentences multiple times, or by adding unrelated software to the prediction.
Both models also tend to be extractive when the augmented text is of a certain length. For instance, if the text is short (20 words), both models will tend to make up summrization that was learned during training by including vulnerability description such as gain access or code execution, even when none of these were mentioned in augmented text.
On the other hand, when the augmented description is too long, the prediction becomes repetitive and hard to understand, although it still covers different portions of the target summary.
One possible solution is to ensure a diversity among the augmented sentences and that no sentence is repeated.
However, this could be expensive, as it requires checking every new candidate paragraph against all  already augmented paragraphs.

\section{Conclusion}
We leverage publicly available resources to enhance and enrich  vulnerabilities description through data augmentation. Our method relies on public databases for collection of text data and pass them through multiple filters to extract relevant text that could contribute to our dataset. We fine-tune two pretrained models that excel in summrization tasks using our curated dataset and report initial and promising result using computational and human metrics. Data curation is a future direction for improving accuracy.

\BfPara{Acknowledgement} This work was supported in part by NRF under grant number 2016K1A1A2912757 and by CyberFlorida's Seed Grant.


\begin{thebibliography}{10}
\providecommand{\url}[1]{\texttt{#1}}
\providecommand{\urlprefix}{URL }
\providecommand{\doi}[1]{https://doi.org/#1}

\bibitem{alabduljabbar2021automated}
Alabduljabbar, A., Abusnaina, A., Meteriz-Yildiran, {\"U}., Mohaisen, D.:
  Automated privacy policy annotation with information highlighting made
  practical using deep representations. In: Proceedings of the 2021 ACM SIGSAC
  Conference on Computer and Communications Security. pp. 2378--2380 (2021)

\bibitem{alabduljabbar2022measuring}
Alabduljabbar, A., Mohaisen, D.: Measuring the privacy dimension of free
  content websites through automated privacy policy analysis and annotation.
  In: Companion Proceedings of the Web Conference (2022)

\bibitem{abs-2006-15074}
Anwar, A., Abusnaina, A., Chen, S., Li, F., Mohaisen, D.: Cleaning the {NVD:}
  comprehensive quality assessment, improvements, and analyses. CoRR
  \textbf{abs/2006.15074} (2020), \url{https://arxiv.org/abs/2006.15074}

\bibitem{anwar2020measuring}
Anwar, A., Khormali, A., Choi, J., Alasmary, H., Choi, S.J., Salem, S., Nyang,
  D., Mohaisen, D.: Measuring the cost of software vulnerabilities. EAI
  Endorsed Transactions on Security and Safety  \textbf{7}(23),  e1--e1 (2020)

\bibitem{AnwarKNM18}
Anwar, A., Khormali, A., Nyang, D., Mohaisen, A.: Understanding the hidden cost
  of software vulnerabilities: Measurements and predictions. In: Beyah, R.,
  Chang, B., Li, Y., Zhu, S. (eds.) Security and Privacy in Communication
  Networks - 14th International Conference, SecureComm 2018, Singapore, August
  8-10, 2018, Proceedings, Part {I}. Lecture Notes of the Institute for
  Computer Sciences, Social Informatics and Telecommunications Engineering,
  vol.~254, pp. 377--395. Springer (2018). \doi{10.1007/978-3-030-01701-9\_21},
  \url{https://doi.org/10.1007/978-3-030-01701-9\_21}

\bibitem{ref_proc9}
Cer, D., Yang, Y., Kong, S.y., Hua, N., Limtiaco, N., John, R.S., Constant, N.,
  Guajardo-Cespedes, M., Yuan, S., Tar, C., et~al.: Universal sentence encoder.
  arXiv preprint arXiv:1803.11175  (2018)

\bibitem{ref_proc5}
Devlin, J., Chang, M.W., Lee, K., Toutanova, K.: Bert: Pre-training of deep
  bidirectional transformers for language understanding. In: 2019 Conference of
  the North American Chapter of the Association for Computational Linguistics
  (2018)

\bibitem{ref_proc1}
Dong, Y., Guo, W., Chen, Y., Xing, X., Zhang, Y., Wang, G.: Towards the
  detection of inconsistencies in public security vulnerability reports. In:
  28th USENIX Security Symposium. pp. 869--885 (2019)

\bibitem{web_ref}
{Help Net Security}: Still relying solely on cve and nvd for vulnerability
  tracking? bad idea.
  \url{https://www.helpnetsecurity.com/2018/02/16/cve-nvd-vulnerability-tracking/}
  (August 2018)

\bibitem{web_ref2}
{Information Security Buzz}: Why critical vulnerabilities do not get reported
  in the cve/nvd databases and how organisations can mitigate the risks.
  \url{https://informationsecuritybuzz.com/articles/why-critical-vulnerabilities/}
  (August 2018)

\bibitem{ref_proc3}
Kanakogi, K., Washizaki, H., Fukazawa, Y., Ogata, S., Okubo, T., Kato, T.,
  Kanuka, H., Hazeyama, A., Yoshioka, N.: Tracing capec attack patterns from
  cve vulnerability information using natural language processing technique.
  In: 54th Hawaii International Conference on System Sciences (2021)

\bibitem{ref_article1}
Kanakogi, K., Washizaki, H., Fukazawa, Y., Ogata, S., Okubo, T., Kato, T.,
  Kanuka, H., Hazeyama, A., Yoshioka, N.: Tracing cve vulnerability information
  to capec attack patterns using natural language processing techniques.
  Information  \textbf{12}(8), ~298 (2021)

\bibitem{kudo2018subword}
Kudo, T.: Subword regularization: Improving neural network translation models
  with multiple subword candidates. arXiv preprint arXiv:1804.10959  (2018)

\bibitem{kudo2018sentencepiece}
Kudo, T., Richardson, J.: Sentencepiece: A simple and language independent
  subword tokenizer and detokenizer for neural text processing. arXiv preprint
  arXiv:1808.06226  (2018)

\bibitem{ref_proc11}
Lewis, M., Liu, Y., Goyal, N., Ghazvininejad, M., Mohamed, A., Levy, O.,
  Stoyanov, V., Zettlemoyer, L.: Bart: Denoising sequence-to-sequence
  pre-training for natural language generation, translation, and comprehension.
  arXiv preprint arXiv:1910.13461  (2019)

\bibitem{mohaisen2015amal}
Mohaisen, A., Alrawi, O., Mohaisen, M.: Amal: high-fidelity, behavior-based
  automated malware analysis and classification. computers \& security
  \textbf{52},  251--266 (2015)

\bibitem{ref_proc12}
Radford, A., Narasimhan, K., Salimans, T., Sutskever, I.: Improving language
  understanding by generative pre-training. OpenAI  (2018)

\bibitem{ref_proc6}
Raffel, C., Shazeer, N., Roberts, A., Lee, K., Narang, S., Matena, M., Zhou,
  Y., Li, W., Liu, P.J.: Exploring the limits of transfer learning with a
  unified text-to-text transformer. arXiv preprint arXiv:1910.10683  (2019)

\bibitem{ref_proc15}
Sennrich, R., Haddow, B., Birch, A.: Neural machine translation of rare words
  with subword units. In: 54th Annual Meeting of the Association for
  Computational Linguistics (2015)

\bibitem{ref_proc8}
Song, K., Tan, X., Qin, T., Lu, J., Liu, T.Y.: Mpnet: Masked and permuted
  pre-training for language understanding. Advances in Neural Information
  Processing Systems  \textbf{33},  16857--16867 (2020)

\bibitem{ref_proc10}
Vaswani, A., Shazeer, N., Parmar, N., Uszkoreit, J., Jones, L., Gomez, A.N.,
  Kaiser, {\L}., Polosukhin, I.: Attention is all you need. Advances in neural
  information processing systems  \textbf{30} (2017)

\bibitem{ref_lncs1}
W{\aa}reus, E., Hell, M.: Automated cpe labeling of cve summaries with machine
  learning. In: International Conference on Detection of Intrusions and
  Malware, and Vulnerability Assessment. pp. 3--22. Springer (2020)

\bibitem{ref_proc7}
Yang, Z., Dai, Z., Yang, Y., Carbonell, J., Salakhutdinov, R., Le~QV, X.:
  generalized autoregressive pretraining for language understanding; 2019.
  Preprint at https://arxiv. org/abs/1906.08237 Accessed June  \textbf{21}
  (2021)

\bibitem{ref_proc13}
Zhang, J., Zhao, Y., Saleh, M., Liu, P.: Pegasus: Pre-training with extracted
  gap-sentences for abstractive summarization. In: International Conference on
  Machine Learning. pp. 11328--11339. PMLR (2020)

\end{thebibliography}
\end{document}